\documentclass[a4paper,12pt]{article}
\pdfoutput=1
\usepackage{graphicx,rotating,amssymb,amsmath,cite}
\usepackage[normalem]{ulem}

\ifx\pdfoutput\undefined
\usepackage[dvips,bookmarks]{hyperref}	
\else
\usepackage{hyperref}	
\fi
\hypersetup{colorlinks,bookmarksopen,bookmarksnumbered,citecolor=rossos,
linkcolor=verde,pdfstartview=FitH,urlcolor=rossos}

\def\hhref#1{\href{http://arxiv.org/abs/#1}{#1}} 

\usepackage{footnote}
\usepackage{booktabs}
\usepackage{multicol}
\usepackage{bold-extra}
\usepackage{color}
\definecolor{rosso}{cmyk}{0,1,1,0.4}
\definecolor{rossos}{cmyk}{0,1,1,0.55}
\definecolor{rossoc}{cmyk}{0,1,1,0.2}
\definecolor{blu}{cmyk}{1,1,0,0.3}
\definecolor{blus}{cmyk}{1,1,0,0.6}
\definecolor{bluc}{cmyk}{1,1,0,0.1}
\definecolor{verde}{cmyk}{0.92,0,0.59,0.25}
\definecolor{verdec}{cmyk}{0.92,0,0.59,0.15}
\definecolor{verdes}{cmyk}{0.92,0,0.59,0.4}

\font\tenrsfs=rsfs10 at 12pt
\font\sevenrsfs=rsfs7
\font\fiversfs=rsfs5
\newfam\rsfsfam
\textfont\rsfsfam=\tenrsfs
\scriptfont\rsfsfam=\sevenrsfs
\scriptscriptfont\rsfsfam=\fiversfs
\def\mathscr#1{{\fam\rsfsfam\relax#1}}

\def\circa#1{\,\raise.3ex\hbox{$#1$\kern-.75em\lower1ex\hbox{$\sim$}}\,}

\newcommand{\beq}{\begin{equation}}
\newcommand{\eeq}{\end{equation}}

\def\circa#1{\,\raise.3ex\hbox{$#1$\kern-.75em\lower1ex\hbox{$\sim$}}\,}
\makeatletter

\def\art{\@ifnextchar[{\eart}{\oart}}
\def\eart[#1]#2#3#4#5#6{{\rm #2}, {#3 #4} {\rm (#6) #5} [{\hhref{#1}}]}
\def\hepart[#1]#2{{\rm #2, \hhref{#1}}}
\newcommand{\oart}[5]{{\rm #1}, {#2 #3} {\rm (#5) #4}}

\newcounter{alphaequation}[equation]
\def\thealphaequation{\theequation\hbox to
0.6em{\hfil\alph{alphaequation}\hfil}}
\def\eqnsystem#1{
\def\@eqnnum{{\rm (\thealphaequation)}}
\def\@@eqncr{\let\@tempa\relax \ifcase\@eqcnt \def\@tempa{& & &} \or
  \def\@tempa{& &}\or \def\@tempa{&}\fi\@tempa
  \if@eqnsw\@eqnnum\refstepcounter{alphaequation}\fi
\global\@eqnswtrue\global\@eqcnt=0\cr}
\refstepcounter{equation} \let\@currentlabel\theequation \def\@tempb{#1}
\ifx\@tempb\empty\else\label{#1}\fi
\refstepcounter{alphaequation}
\let\@currentlabel\thealphaequation
\global\@eqnswtrue\global\@eqcnt=0 \tabskip\@centering\let\\=\@eqncr
$$\halign to \displaywidth\bgroup \@eqnsel\hskip\@centering
$\displaystyle\tabskip\z@{##}$&\global\@eqcnt\@ne
\hskip2\arraycolsep\hfil${##}$\hfil& \global\@eqcnt\tw@\hskip2\arraycolsep
$\displaystyle\tabskip\z@{##}$\hfil
\tabskip\@centering&\llap{##}\tabskip\z@\cr}
\def\endeqnsystem{\@@eqncr\egroup$$\global\@ignoretrue} \makeatother

\begin{document}
\begin{flushright}
\footnotesize
{FERMILAB-PUB-23-108-T}
\end{flushright}
\color{black}

\begin{center}
{\Huge\bf Putting all the X in one basket:\\[3mm] Updated X-ray constraints on \\[3mm] sub-GeV Dark Matter}

\medskip
\bigskip\color{black}\vspace{0.6cm}

{
{\large\bf Marco Cirelli}\ $^a$,
{\large\bf Nicolao Fornengo}\ $^b$,\\[3mm]
{\large\bf Jordan Koechler}\ $^{a}$\,\footnote{Now at: \href{https://www.to.infn.it/}{Istituto Nazionale di Fisica Nucleare}, Sezione di Torino, Via P. Giuria 1, I-10125 Torino, Italy.},
{\large\bf Elena Pinetti}\ $^{c,d}$\,\footnote{Now at: \href{https://www.simonsfoundation.org/flatiron/center-for-computational-astrophysics/}{Center for Computational Astrophysics}, Flatiron Institute, New York, NY 10010, USA.}
{\large\bf Brandon M. Roach}\ $^{e}$\,\footnote{Now at: \href{https://kavlicosmo.uchicago.edu}{Kavli Institute for Cosmological Physics}, University of Chicago, Chicago, IL 60637, USA.}
}
\\[7mm]
{\it $^a$ \href{http://www.lpthe.jussieu.fr/spip/index.php}{Laboratoire de Physique Th\'eorique et Hautes Energies (LPTHE)},\\ UMR 7589 CNRS \& Sorbonne University, 4 Place Jussieu, F-75252, Paris, France}\\[3mm]
{\it $^b$ \href{http://www.df.unito.it/do/home.pl}{Dipartimento di Fisica}, Universit\`a di Torino \& INFN, Sezione\ di Torino,\\ via P. Giuria 1, I-10125 Torino, Italy}\\[3mm]
{\it $^c$ \href{https://astro.fnal.gov/science/theory}{Theoretical Astrophysics Department}, Fermi National Accelerator Laboratory,\\ Batavia, Illinois 60510, USA}\\[3mm]
{\it $^d$ \href{https://kavlicosmo.uchicago.edu}{Kavli Institute for Cosmological Physics}, University of Chicago,\\ Chicago, IL 60637, USA}\\[3mm]
{\it $^e$ \href{https://physics.mit.edu/}{Department of Physics}, Massachusetts Institute of Technology,\\ Cambridge, Massachusetts 02139, USA}\\
\end{center}

\bigskip

\centerline{\large\bf Abstract}
\begin{quote}
\large
Sub-GeV dark matter particles can annihilate or decay producing $e^\pm$ pairs which upscatter the low-energy photon fields in the Galaxy and generate an $X$-ray emission (via the Inverse Compton effect). 
Using $X$-ray data from {\sc Xmm-Newton}, {\sc Integral}, {\sc NuStar} and {\sc Suzaku}, we derive new constraints on this class of dark matter (DM).  For annihilating DM, they are significant for $m_{\rm DM} \gtrsim 100 \ {\rm MeV}$, and dominant if DM is $p$-wave annihilating. For decaying DM, they are the most stringent to date in the range $m_{\rm DM} \simeq 400 \ {\rm MeV} - 3 \ {\rm GeV}$.
\end{quote}

\newpage

\tableofcontents

\section{Introduction}
\label{sec:introduction}

The possibility that Dark Matter (DM) consists of a {\em light} particle, where by light we mean that its mass $m_{\rm DM}$ is in the range  $m_{\rm DM} \simeq 1 \ {\rm MeV}$ - few GeV, has received significant attention recently. This is in part a reaction to the lack of convincing signals of the long sought-after weak-scale DM in current experiments \cite{Schumann:2019eaa,Cirelli:2015gux,Gaskins:2016cha,Hooper:2018kfv,Buchmueller:2017qhf,Kahlhoefer:2017dnp}, and in part a consequence of the emerging of motivated sub-GeV DM theoretical models \cite{Knapen:2017xzo,Boehm:2002yz,Boehm:2003hm,Fayet:2007ua,Boehm:2003bt,Ahn:2005ck,Boehm:2006mi,Ema:2020fit,Hochberg:2014dra,Boddy:2014yra,Hochberg:2014kqa,Choi:2017zww,Berlin:2018tvf,DAgnolo:2015ujb,Falkowski:2011xh,Lin:2011gj,Hooper:2007tu,Bertuzzo:2017lwt,Darme:2017glc,Katz:2020ywn}. 
For review of all these aspects, the interested reader can refer to our previous work~\cite{Cirelli:2020bpc}. 
Here it suffices to recall that the detection of light DM is more challenging than weak-scale DM. 
In indirect detection, the main obstacle consists in the so-called `MeV gap', i.e.~the fact that no recent high-sensitivity data exist in the $\gamma$-ray energy window $\sim$100 keV - 100 MeV, corresponding to the  interval where the signals from light DM particle annihilation or decay are expected.

A novel technique introduced in \cite{Cirelli:2020bpc} allows to circumvent this problem. The idea is to focus on {\em secondary} emissions from DM, and in particular on the Inverse Compton Scattering (ICS) process. Namely, DM annihilations or decays in the galaxy produce electrons and positrons which can upscatter the low energy photons of the ambient light (whose main components are the optical light from stars, the infrared light from dust and the CMB) and produce hard $X$-rays with typical keV energy.  As a result, one can leverage on the abundant  data in $X$-ray keV observations, rather than the scarce MeV  experiments, in order to test sub-GeV DM. 

In \cite{Cirelli:2020bpc} it was shown that the method is powerful. Using data from a large region of the inner galaxy observed by the {\sc Integral/Spi} spectrometer, \cite{Cirelli:2020bpc} was able to obtain stringent constraints on annihilating DM in the mass range 1 MeV to 5 GeV. 
Following up on that work, we embark here in a systematic analysis of the available datasets in $X$-rays in order to assess their full constraining power on light DM, along the lines of the strategy described above. In addition, we consider both the case of annihilating and decaying DM. 
The main challenge in this endeavour consists in the fact that often $X$-ray experiments are not focused on wide surveys of the sky, but instead on point sources or small areas of observations. 
Still, we will show that by adopting different observational data, referring to different energy ranges and angular positions in the sky, we can somewhat improve the bounds on both annihilating and decaying DM. 

\medskip

{\bf Note added in v3.} 
In 2025, Ref.~\cite{Balaji:2025afr} realized that the v1 and v2 of our present work (as well as \cite{DelaTorreLuque:2023olp}), which were using for continuum searches the {\sc Xmm-Newton} data provided in \cite{Dessert:2018qih,Foster:2021ngm}, had missed a normalization factor in the {\tt github} repository\footnote{See on \url{https://github.com/bsafdi/XMM_BSO_DATA}.}. A data-loading tutorial showing the procedure to use the data is now available on that repository and was not present at the time of our initial analysis. Properly including this normalization `weight' significantly modifies the results concerning {\sc Xmm-Newton} that were presented in v1 and v2 of this paper. See below for more details.

\medskip

The  paper is organized as follows:  in sec.~\ref{sec:flux} we briefly recall the formalism and the relevant quantities necessary for computing prompt and ICS $X$-ray emissions from light DM annihilations and decays; in sec.~\ref{sec:data} we detail the dataset that we use; in sec.~\ref{sec:results} we present our analysis and the main results, and in sec.~\ref{sec:comparison} we compare with related studies. 
Finally, in sec.~\ref{sec:conclusions} we draw our conclusions.


\section{X-rays from DM annihilations and decays}
\label{sec:flux}

In this section we discuss the basic formalism for $X$-ray production from DM annihilations and decays. 
Here we only recall the main ingredients and focus in particular on the novelties of the present paper. The interested reader can refer to \cite{Cirelli:2020bpc} for the detailed and complete formalism. 

\medskip

We are dealing with DM lighter than a few GeV, hence we consider only three annihilation or decay channels:
\begin{align}
{\rm DM \, (DM)} &\to e^+e^-,  \label{eq:ee}\\
{\rm DM \, (DM)} &\to \mu^+\mu^-, \label{eq:mumu}\\
{\rm DM \, (DM)} &\to \pi^+\pi^-, \label{eq:pipi}
\end{align}
which are kinematically open whenever $m_{\rm DM} > m_i$ (annihilations) or $m_{\rm DM} > 2 m_i$ (decays), with $i=e,\mu,\pi$. 
We consider the channels one at a time independently although of course, in specific models, DM could annihilate or decay in a combination of modes that can also include other light hadronic or mesonic resonances. A more thorough model-dependent study can be done by computing photon energy spectra using available numerical codes \cite{Coogan:2019qpu, Coogan:2022cdd} and applying them to our study. We leave this analysis to a future work.

Given a fixed channel, the total flux of photons is given by the sum of two contributions: (i) the prompt emission from the charged particles in the final state and (ii) the secondary emission of photons produced via ICS by the energetic $e^\pm$ originating from DM annihilations or decays. 
In turn, the prompt emission consists of Final State Radiation (FSR) from the charged leptons or pions in the final state, and of radiative decays (Rad) which occur whenever muons or pions undergo a decay with an extra photon involved ($\mu \to e\nu_e\nu_\mu\gamma$, $\pi \to l\nu_l\gamma$, with $l=e,\mu$ $-$ this notation comprises particles and antiparticles and can be adapted in an obvious way). 

\medskip

The differential flux of the prompt emissions is readily computed as the usual integral of the emissions along the line of sight (l.o.s.) in a given direction $\theta$, the angle with respect to the direction to the Galactic Center (GC), and parameterized by $s$:
\begin{equation}
\label{eq:promptflux}
   \frac{d\Phi_{{\rm prompt} \, \gamma}}{dE_\gamma \, d\Omega}=\frac{1}{4\pi}\frac{dN_{{\rm prompt} \, \gamma}}{dE_\gamma}
  \times \left\{
\begin{array}{ll}
	\displaystyle  
        \frac{\langle \sigma v\rangle}{2} \int_\text{l.o.s.}ds\,\left(\frac{\rho_\text{DM}(r(s,\theta))}{m_\text{DM}}\right)^2 &\text{(annihilation)}\\
	\\
        \displaystyle
        \ \; \Gamma \ \; \int_\text{l.o.s.}ds\,\left(\frac{\rho_\text{DM}(r(s,\theta))}{m_\text{DM}}\right) &\text{(decay)}
    \end{array}
\right.    
.
\end{equation}
Here $\langle \sigma v\rangle$ and $\Gamma$ represent the thermally averaged DM annihilation cross section and the DM decay rate, respectively. The photon spectra $dN_{{\rm prompt} \, \gamma}/dE_\gamma$, where prompt = FSR or Rad, are given by the lengthy but straightforward expressions provided in \cite{Cirelli:2020bpc}. The DM density profile $\rho_\text{DM}$ in the Milky Way is assumed to be a standard Navarro-Frenk-White (NFW) profile \cite{Navarro:1995iw} with the parameters specified in \cite{Cirelli:2020bpc,Cirelli:2010xx}, and we will investigate the impact of modifying this choice in sec.~\ref{sec:results}. 

\medskip

The differential flux of the ICS emission is given by
\beq
\label{eq:ICSflux}
\frac{d\Phi_{{\rm IC}\gamma}}{dE_\gamma \, d\Omega} = \frac 1{E_\gamma} \int_{\rm l.o.s.} ds\, \frac{j(E_\gamma, s, b, \ell)}{4\pi},
\eeq
where the set of coordinates $(s,b,\ell)$ indicates a unique position in the galactic halo, with $(b, \ell)$ the galactic latitude and longitude. The emissivity $j$ at a given point is the convolution of the ICS power $\mathcal{P}_{\rm IC}$ with the differential number density $dn_{e^\pm}/dE_e$ of emitting electrons and positrons present at that point:
\beq
\label{Rademissivity}
j(E_\gamma, s, b, \ell)=2\int_{m_e}^{m_{\rm DM}(/2)}dE_e\ \mathcal{P}_{\rm IC}(E_\gamma,E_e, s, b, \ell)\ \frac{dn_{e^\pm}}{dE_e}(E_e,s,b,\ell).
\eeq
Note that the integration over the $e^\pm$ energy $E_e$ runs from the electron mass $m_e$ to the maximal possible value, corresponding to the DM rest mass for annihilations and half of it for decays. The ICS differential power, i.e. the power per photon energy
\begin{equation}
    \mathcal{P}_{\rm IC}(E_\gamma,E_e,s, b, \ell)=E_\gamma\int_0^1 dy\,n_\gamma(E_\gamma^0(y),s, b, \ell)\,\sigma_{\rm IC}(y,E_e)
\end{equation}
 includes the density of photons per unit energy $n_\gamma$, on which the $e^\pm$ scatter with cross section $\sigma_{\rm IC}$, the Klein-Nishina cross section in the Thomson limit ($E_e \gg E_\gamma^0$), where $E_\gamma^0$ denotes the initial photon energy, $y=E_\gamma/(4\gamma_e E_\gamma^0)$ and $\gamma_e=E_e/m_e$. The $e^\pm$ number density is determined, in the so-called `on-the-spot approximation', as
\beq
\frac{dn_{e^\pm}}{dE_e}(E_e, s, b, \ell) = \frac1{b_{\rm tot}(E_e, s, b, \ell)} \times  
   \left\{
\begin{array}{ll}
 \displaystyle  
 \int_{E_e}^{m_{\rm DM}}d\tilde E_e \frac{\langle \sigma v\rangle}{2} \left( \frac{\rho(s,b,\ell)}{m_{\rm DM}} \right)^2 \frac{dN_{e^\pm}}{d \tilde E_e} &\text{(annihilation)}\vspace{2mm} \\ 
 \displaystyle
 \int_{E_e}^{m_{\rm DM/2}}d\tilde E_e  \ \; \Gamma \  \left( \frac{\rho(s,b,\ell)}{m_{\rm DM}} \right) \ \frac{dN_{e^\pm}}{d \tilde E_e} &\text{(decay)}
     \end{array}
\right.    
.
\eeq
Here $b_{\rm tot}$ is the energy loss function, which takes into account all the energy loss processes that the $e^\pm$ suffer in the local Galactic environment in which they are injected. The $e^\pm$ spectra from DM annihilations or decays in the different channels are computed following \cite{Cirelli:2020bpc}: for the $e^+e^-$ channel the spectrum consists simply in a monochromatic line with $E_e = m_{\rm DM}$; for the $\mu^+\mu^-$ channel it consists of a boosted Michel spectrum from muon decay; for the $\pi^+\pi^-$ channel it consists of a doubly boosted Michel spectrum. 

With these ingredients, we are able to compute the full spectrum of photons from DM annihilations and decays. 
As a final step, one integrates the contributions in eqs.~(\ref{eq:promptflux}) and (\ref{eq:ICSflux}) over the selected region of observation, identified by intervals in $b$ and $\ell$:
\beq
\label{eq:totaldiffflux}
\frac{d\Phi_{{\rm DM}\gamma}}{dE_\gamma}= \int_{b_{\rm min}}^{b_{\rm max}} \int_{\ell_{\rm min}}^{\ell_{\rm max}} db \, d\ell \, \cos b \ \left( \frac{d\Phi_{{\rm prompt}\gamma}}{dE_\gamma \, d\Omega} + \frac{d\Phi_{{\rm IC}\gamma}}{dE_\gamma\, d\Omega} \right).
\eeq

\medskip

\begin{figure}[!t]
\begin{center}
\includegraphics[width= 0.48 \textwidth]{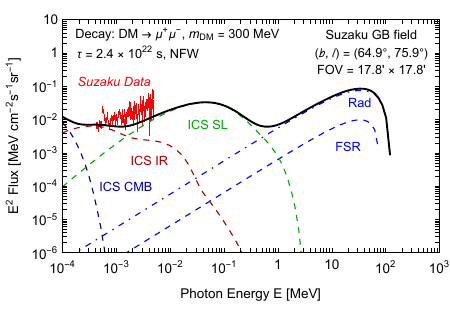} \quad
\includegraphics[width= 0.48 \textwidth]{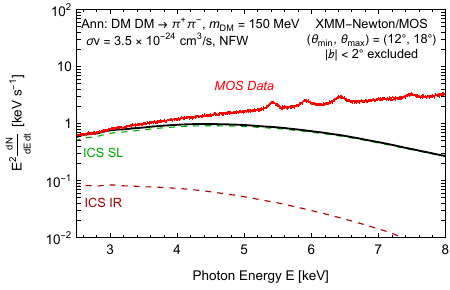} \\
\includegraphics[width= 0.48 \textwidth]{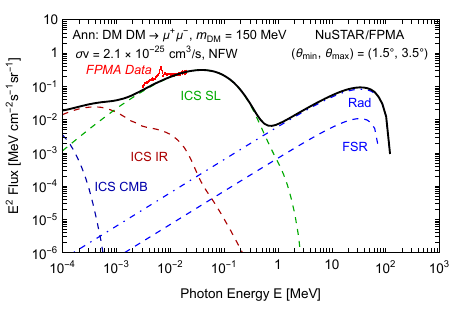} \quad
\includegraphics[width= 0.48 \textwidth]{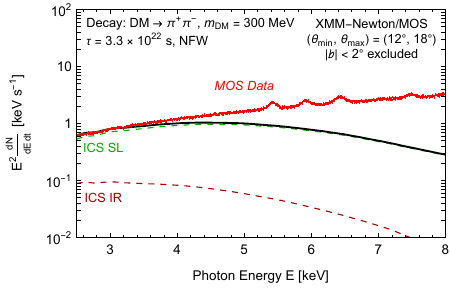} \\
\includegraphics[width= 0.48 \textwidth]{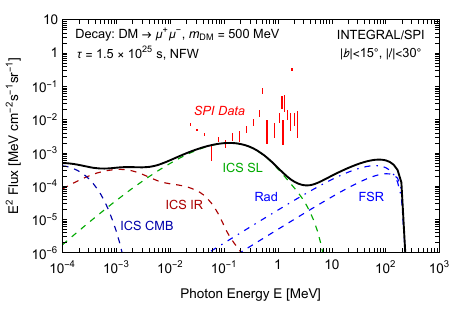} \quad
\includegraphics[width= 0.48 \textwidth]{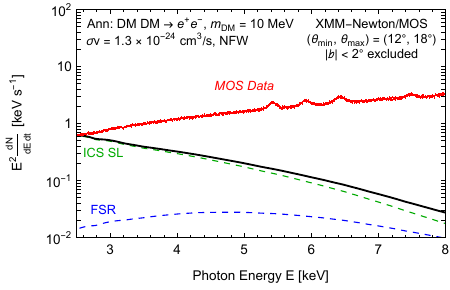}
\caption{\em \small \label{fig:fluxes} Illustration of some fluxes of hard $X$-rays from DM annihilation or decay, compared to the different datasets adopted in our analysis. In each panel we indicate the DM specifications (annihilation or decay channel, mass, annihilation cross section or decay rate, galactic distribution $-$always NFW) and the characteristics of the considered region of observation.}
\end{center}
\end{figure}

Fig.~\ref{fig:fluxes} illustrates a few examples of the total flux, compared to the datasets that we considered in our analysis. Such datasets are discussed in the next section. 


\section{Datasets and analysis}
\label{sec:data}

\begin{figure}[t]
\begin{center}
\includegraphics[width= 0.80 \textwidth]{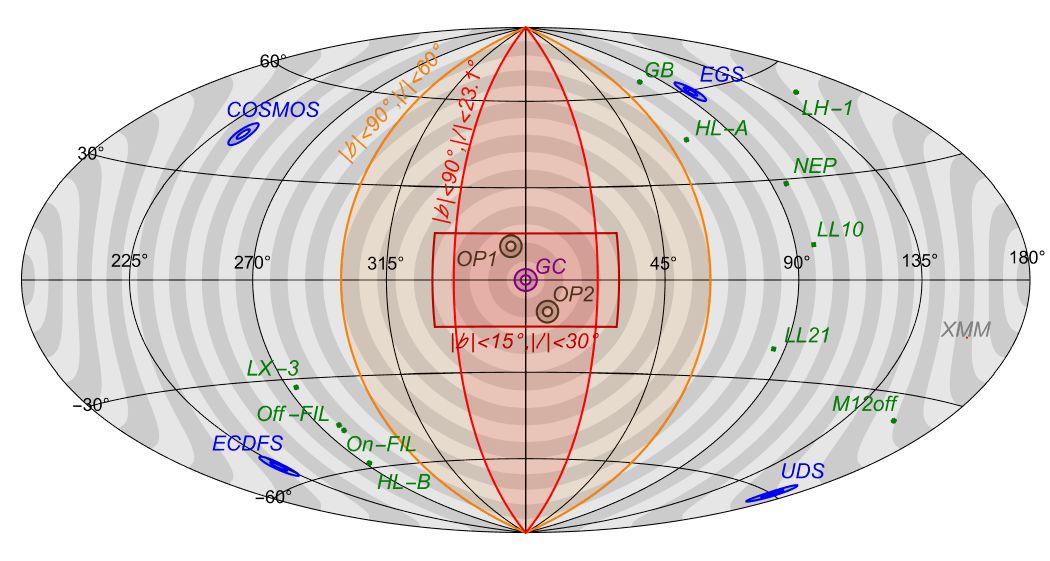}
\caption{\em \small \label{fig:galaxymap} Chart of the Galaxy in galactic coordinates with the location of the datasets we use. The three regions of observation relevant for the {\sc Integral} datasets are represented in orange and red. The four {\sc NuStar} Blank-Sky fields (COSMOS, EGS, ECDFS and UDS) are in blue, the {\sc NuStar} Galactic Center (GC) and Off-Plane (OP1 \& OP2) are in purple and dark brown, respectively. {\sc Xmm-Newton} rings are drawn in shades of grey and the eleven {\sc Suzaku} fields in green. The fields are to scale.
} 
\end{center}
\end{figure}

In this study we focus on the $X$-ray emission of the Milky Way galaxy and we exploit the datasets listed below. The locations of the respective regions of interest on the galactic sky are depicted for illustration in fig.~\ref{fig:galaxymap}.

\begin{itemize}
\item {\color{blue} {\sc Integral}}. The data are reported in \cite{Bouchet:2011fn}, which follows previous work in \cite{Bouchet:2008rp,Bouchet:2005ys}. These datasets were used in our previous paper \cite{Cirelli:2020bpc}. 
The data were collected by the {\sc Spi} $X$-ray spectrometer onboard the {\sc Integral}  satellite, in the period 2003$-$2009, corresponding to a significant total exposure of about $100$ Ms, and cover a range in energy between 20 keV and a few MeV. They are provided either in the form of a spectrum of the total diffuse flux in a rectangular region of observation centered around the GC ($|b|<15^\circ, |\ell|<30^\circ$, figs.~6 and 7 in \cite{Bouchet:2011fn}) or in the form of an angular flux in latitude and longitude bins, in 5 energy bands (27$-$49 keV, 49$-$90 keV, 100$-$200 keV, 200$-$600 keV and 600$-$1800 keV) (figs.~4 and 5 in \cite{Bouchet:2011fn}). 
As in \cite{Cirelli:2020bpc}, we use the angular flux in latitude bins only, from which we cut out the Galactic Plane (GP). The longitude window is $|\ell|<23.1^\circ$ for the first four energy bands and $|\ell|<60^\circ$ for the fifth one.

\item {\color{blue} {\sc NuStar} Blank-Sky} fields. These data are presented in \cite{Krivonos:2020qvl}, which aims at measuring the cosmic X-ray background (CXB) in the $3-20$ keV energy band. 
The data are collected from the {\sc NuStar} extragalactic survey program, which includes a number of fields with different sky coverage and exposure times, among which there are the COSMOS, EGS, ECDFS, UDS that we use. 
These are the same fields used in \cite{Roach:2022lgo}, although in another context (namely, to probe sterile neutrino DM). 
The actual areas of observation have a complex shape: they consist of two partly overlapping `Pac-Man$^{\scriptscriptstyle \rm TM}$-like' regions located around the nominal pointing center of the field, with uneven coverage (see e.g.~fig.~4 in \cite{Perez:2016tcq}). We choose to approximate each of them as a square annulus of inner size $1.5^\circ$ and outer size $3.5^\circ$. This approximation is justified by the fact that the DM emissivity in those relatively small regions varies little, thus we can adopt a simpler geometrical area.
The nominal exposure is of about 7 Ms.

\item {\color{blue} {\sc NuStar} Galactic Center (GC)} region. The data are provided in \cite{Mori:2015vba,Hong:2016qjq} and are the same used in \cite{Perez:2016tcq} in another context (namely, to probe sterile neutrino DM). 
The shape of the areas of observation is the same as in the previous item: we just model it here as an annulus of inner radius $1.5^\circ$ and outer radius $3.5^\circ$.
We use the data provided in fig.~5 of \cite{Perez:2016tcq}, restricting at $E_\gamma \le 20$ keV because the instrumental background becomes dominant for higher energies\footnote{We should note, however, that these spectra (even for $E_\gamma < 20$ keV) include a small contribution from internal detector background, 
which we do not model nor subtract. This implies that our bounds are derived from a nominal flux which is sligthtly larger that the true astrophysical emission: thus, the derived DM limits are conservative compared to the approach where the full background is modelled.}. Since this emission originated from regions close to the GC, it is subject to attenuation upon the dense interstellar medium. However, using a column density of $1 \times 10^{22}/{\rm cm}^2$\cite{HI4PI} and the cross sections tabulated in \cite{Wilms:2000ez}, we find that such attenuation is at most $\sim$10\% at $E_\gamma = 3$ keV and quickly diminishes at higher energies, hence it is negligible for our purposes.  

\item {\color{blue} {\sc NuStar} Off-Plane (OP)} Faint-Sky Observations. The data are presented and used in \cite{Roach:2019ctw}. They correspond to the observation of two annuli, with shapes equivalent to those described for the previous datasets (which we model as in the `Blank-Sky' case), located about 10 degrees above and below the GP. The total exposure time amounts to about 100 ks. 
The emission in these regions is understood to be essentially CXB only, since the galactic component is estimated to be negligible. In particular, the Galactic ridge emission (GRXE)\footnote{The GRXE mostly comes from accreting compact objects, mainly white dwarfs. More specifically, it is believed to be produced in the accretion streams of magnetic cataclysmic variable stars, plus a 6.4 keV Fe I line. The interested reader can find more information in \cite{Roach:2019ctw} and \cite{Krivonos:2020qvl}.} is expected to be small, since it falls off rapidly with increasing latitude, due to the lower stellar density. 
Hence, we use the same data as the {\sc NuStar} Blank-Sky fields, but with error bars scaled up by a factor $\sqrt{7 \, {\rm Ms}/100 \, {\rm ks}} = 8.4$ to account for the shorter exposure time. 
We stress that, given the weak constraining power that {\sc NuStar} turns out to provide (as we will discuss in the following section), these approximations are sufficient for our purposes. As a side remark, note that the {\sc NuStar} data we use were collected by the FPMA and FPMB detectors on board of the satellite. Because the photon spectra measured by the two detectors are similar, the computed constraints have only a negligible difference, thus we only show the results using the FMPA detector.

\item {\color{blue} {\sc Xmm-Newton}} observations. The data are used in \cite{Dessert:2018qih,Foster:2021ngm} to search for decaying sterile neutrino DM. In particular, the data are provided in a very convenient form, which we use extensively\footnote{See on \url{https://github.com/bsafdi/XMM_BSO_DATA}.}. 
They correspond to observations with the two cameras (called MOS and PN) onboard the {\sc Xmm-Newton} satellite, over an extensive period of about 18 years, from the launch of the telescope (in late 1999) to September 2018.
After the removal of point sources, the data are combined into 30 concentric rings of width 6 degrees as measured in angular distance from the GC. 
Importantly, note that the data provided by \cite{Dessert:2018qih,Foster:2021ngm} are normalized by the exposure-weighted average solid angle, which is orders of magnitude smaller than the geometrical one. This weight needs to be correctly included in order not to overestimate the flux. 
A slice of $|b|\le 2^\circ$ is removed, i.e.~the GP is masked. 
The energy range initially covers 2 eV to 20 keV, however we restrict it as prescribed in \cite{Foster:2021ngm} to avoid the dominant instrumental background. The final energy range is therefore 2.5 to 8 keV for MOS and 2.5 to 7 keV for PN. Response matrices for both instruments are also provided.

\item {\color{blue} {\sc Suzaku}} high-latitude fields. The data are provided in \cite{Yoshino:2009kv}, which focuses on measuring the soft diffuse X-ray emission from several small fields located at large galactic longitudes ($65^\circ < \ell < 295^\circ$) and observed for a period of a few days each between 2006 and 2008, using the backside illuminated CCD (BI CCD) of the {\sc Xis} spectrometer on board of the {\sc Suzaku} satellite. We use the data\footnote{The data are shown in figs.~2 and 5 of \cite{Yoshino:2009kv} and we obtained in digital form from M.~Kazuhisa, private communication. The NEP field combines the data from NEP1 and NEP2. We could not obtain the data for the LH-2 field, which we therefore neglect.}
from the 11 fields denoted as: GB, HL-B, LH-1, Off-FIL, On-FIL, HL-A, M12off, LX-3, NEP, LL21 and LL10. We refer to table 1 of \cite{Yoshino:2009kv} for the details of the regions (coordinates, exposures and the original references). We do not consider the R1 and R2 fields, which include bright point sources. From the data, the point sources and the X-ray emission induced by the solar wind proton flux have been carefully removed by the {\sc Suzaku} collaboration. The energy range is 0.4$-$5 keV for all fields, and the typical exposures vary between 16 and 60 ks. The effective area of the experiment in the range of interest roughly equals 100 to 300 cm$^2$. However, we use the detailed published determination (see below).

\end{itemize}

In order to derive the constraints, we first compute the total photon flux from DM annihilation/decay, for each channel considered in Eqs.(\ref{eq:ee}--\ref{eq:pipi}) and region of interest. For the {\sc Integral/Spi} dataset, we compute the photon flux for each latitude bin and energy band. For the remaining datasets we compute the photon flux for each energy bin. Then we correct some of the predicted flux in order to take into account instrumental features:
\begin{itemize}
\item[$\circ$] For each ring of the {\sc Xmm-Newton} dataset, we convolve the photon energy spectrum with the instrumental response function as prescribed in \cite{Kaastra:2016qwt}. Given a specific ring, where $\left(d\Phi_{{\rm DM}\gamma}/dE_\gamma\right)_j=\left(dN_{{\rm DM}\gamma}/dE_\gamma \, dA \, dt\right)_j$ is our predicted DM spectrum in the input energy bin  $j$, the discrete convolution with the instrument response is 
$\left(dN_{{\rm DM}\gamma}/dE_\gamma \, dt\right)_i=\sum_jR_{ij}\left(dN_{{\rm DM}\gamma}/dE_\gamma \, dA\, dt \right)_j$ 
in the output energy bin $i$, where $R_{ij}$ is the instrument response matrix.\footnote{Here by input and output we mean the predicted flux before and after the convolution with the instrumental response matrix, respectively.} The matrices are different for each ring and take into account the effective area of the instrument (in units of cm$^2$).
\item[$\circ$] For the {\sc Suzaku} dataset, we multiply the calculated photon energy spectrum by the {\sc Xis} effective area function as provided on the {\sc Nasa} archives~\footnote{See \url{https://heasarc.gsfc.nasa.gov/docs/suzaku/gallery/performance/xis_area.html}.}. We use the function for the BI CCD.
\end{itemize}

We infer the constraints for each dataset separately via the test statistic:
\beq
\chi_>^2 = \sum_i \left(\frac{{\rm max}[\Phi_{{\rm DM}\gamma,i}(p, m_{\rm DM})-\phi_i,0]}{\sigma_i}\right)^2,
\label{eq:chi2}
\eeq
where $p = \langle \sigma v \rangle$ or $\Gamma$, $\Phi_{{\rm DM}\gamma,i}$ is the predicted photon flux from DM annihilation/decay\footnote{For {\sc Xmm-Newton} and {\sc Suzaku} the flux is actually replaced by the rate of photons per second per keV, the quantity provided by the experiment. For all the other experiments, we use the proper flux.} at the energy (or latitude for {\sc Integral}) bin $i$, $\phi_i$ is the observed flux and $\sigma_i$ its uncertainty. We then impose a $2\sigma$ bound on the parameter $p$ (for each value of $m_{\rm DM}$) whenever we obtain $\chi_>^2=4$.
This procedure means, in particular, that we directly compare the DM prediction with the data, without including any $X$-ray astrophysical background. Including an astrophysical background would in most cases reduce the room for the DM flux and therefore strengthen the constraints. Our procedure thus allows us to derive  conservative bounds. 
In the next section we discuss the obtained constraints.


\section{Results and discussion}
\label{sec:results}

\begin{figure}[t]
\begin{center}
\includegraphics[width= 0.48 \textwidth]{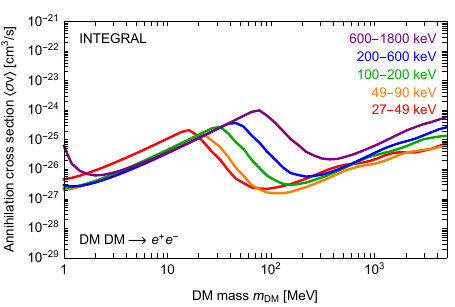} \quad
\includegraphics[width= 0.48 \textwidth]{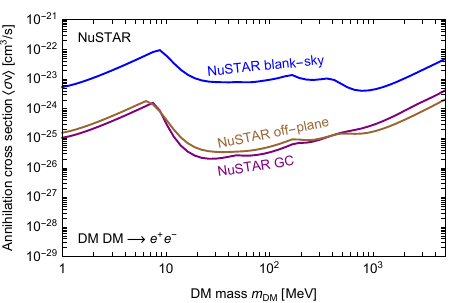} \\
\includegraphics[width= 0.48 \textwidth]{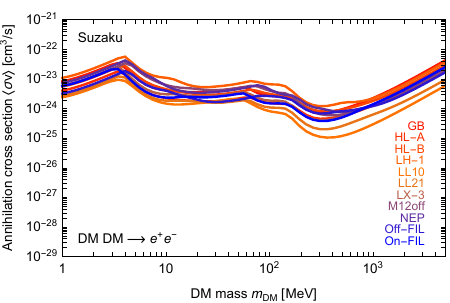} \quad
\includegraphics[width= 0.48 \textwidth]{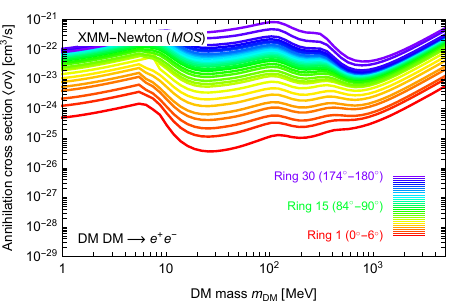}
\caption{\em \small \label{fig:PartialResultsAnn} Conservative constraints on annihilating DM from the different portions of the datasets that we consider. Top left panel: constraints from the different energy bands of the {\sc Integral} dataset (different colors). Top right panel: constraints from the three different regions of observation that we use in the {\sc NuStar} dataset. Bottom left panel: constraints from the eleven different fields of the {\sc Suzaku} dataset (distinguished by the different colors as in the legend). Bottom right panel: constraints from the thirty rings of the {\sc Xmm-Newton} data (distinguished by the different colors as in the legend), for the MOS camera for definiteness.}
\end{center}
\end{figure}

\begin{figure}[t]
\begin{center}
\includegraphics[width= 0.48 \textwidth]{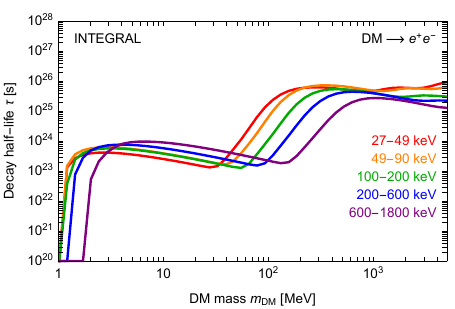} \quad
\includegraphics[width= 0.48 \textwidth]{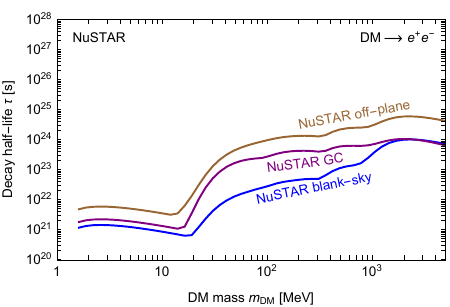} \\
\includegraphics[width= 0.48 \textwidth]{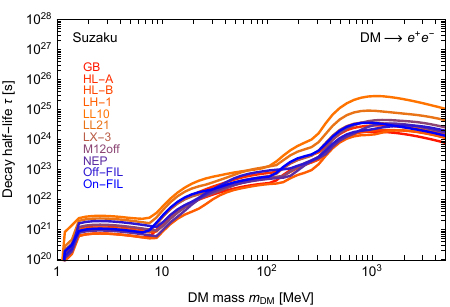} \quad
\includegraphics[width= 0.48 \textwidth]{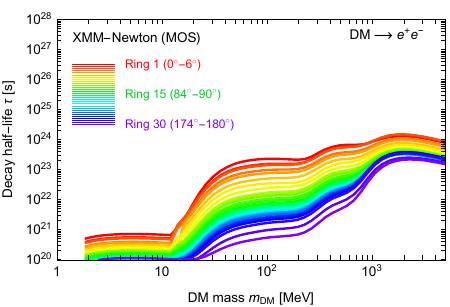}
\caption{\em \small \label{fig:PartialResultsDec} Same as in fig.~\ref{fig:PartialResultsAnn} but for decaying DM.}
\end{center}
\end{figure}

We start by presenting, in fig.~\ref{fig:PartialResultsAnn} for the annihilation case and in fig.~\ref{fig:PartialResultsDec} for the decay case, the conservative constraints obtained from each experiment for each portion of the dataset (either observation subfield or energy band). In each case the bounds are derived using the criterion in eq.~(\ref{eq:chi2}). We focus here for definiteness on the DM (DM) $\to e^+e^-$ channel. 

\medskip

In the top left panels we show the {\sc Integral} bounds imposed by each energy band separately (for the annihilation case, this figure reproduces the analogous one in \cite{Cirelli:2020bpc}). The characteristic shape of the curves is motivated as follows: in the region of large DM masses a strong bound occurs because the ICS flux is constrained by the data points, as shown in the lower left panel of fig.~\ref{fig:fluxes}; the prompt emission is  instead responsible for the bound on small DM masses. In the intermediate mass range the bound is weaker because the data fall in the trough of the characteristic `double hump' shape of the prompt+ICS spectra.
Note that the kink between large and small masses moves to larger DM masses for the higher energy bands and to lower masses for the lower energy bins. This is due to the fact that the DM spectrum shifts to the left with decreasing $m_{\rm DM}$. Overall, given the configuration of the data points and the DM spectra, we find that the low energy bands are more constraining for large masses while high energy bins are more constraining for small masses. 

\medskip

In the top right panel we show the bounds imposed by each {\sc NuStar} dataset separately. The shape of the constraints is analogous to that of {\sc Integral}, with the kink occurring at smaller masses ($m_{\rm DM} \simeq$ 10 MeV) since the {\sc NuStar} data cover lower energies. The limits from the GC region and the Off-Plane fields are more constraining, while those from the Blank-Sky fields are weaker. 
In absolute terms, the {\sc NuStar} results are weaker with respect to the {\sc Integral} ones for the following reasons. For the {\sc NuStar} Blank-Sky case, the fields are at very high latitudes, where the galactic DM emission is small. 
For the {\sc NuStar} GC case, the main component of the measured flux is understood to be the GRXE \cite{Roach:2019ctw}, and the DM flux has to compete with this sizeable foreground: for decaying DM, the DM flux is overwhelmed by the GRXE; for annihilating DM, the DM flux is boosted by the square of the large DM density in the central regions and hence better bounds occur.
The Off-Plane case offers competitive limits overall because, as discussed above, the regions of observation are located enough far away from the plane that the GRXE has decreased and hence the DM contribution can emerge. 

\medskip

\begin{figure}[!t]
\begin{center}
\includegraphics[width= 0.48 \textwidth]{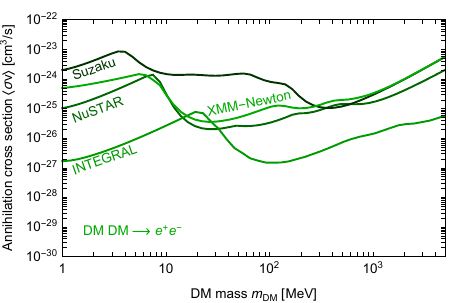} \quad
\includegraphics[width= 0.48 \textwidth]{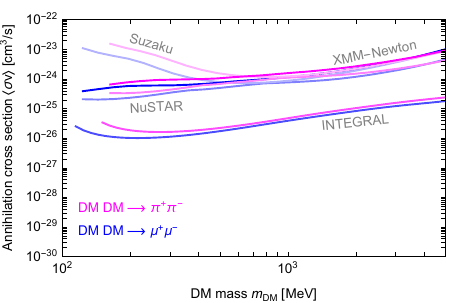} 
\caption{\em \small \label{fig:ResultsAnn} Summary of our conservative constraints on annihilating DM from each experiment and for all channels. The left panel refers to the $e^+e^-$ annihilation channel (green lines), while the right plot to the $\pi^+\pi^-$ (magenta) and $\mu^+\mu^-$ (blue) channels. From top (least constraining) to bottom (most constraining), the experiments are roughly ordered as {\sc Suzaku}, {\sc Xmm-Newton}, {\sc NuStar} and {\sc Integral}.}
\bigskip
\includegraphics[width= 0.48 \textwidth]{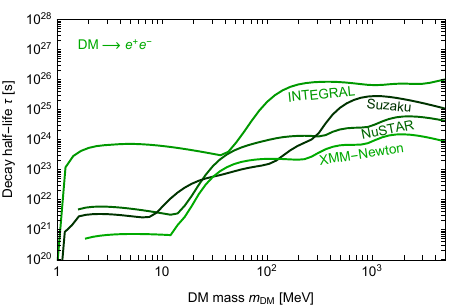} \quad
\includegraphics[width= 0.48 \textwidth]{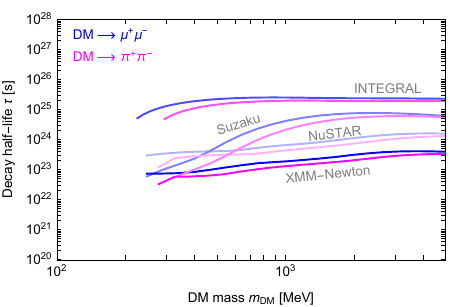} 
\caption{\em \small \label{fig:ResultsDec} Summary of our conservative constraints on decaying DM from each experiment and for all channels. The ordering, now inverted as bottom (least constraining) to top (most constraining), is similar to fig.~\ref{fig:ResultsAnn}.}
\end{center}
\end{figure}

In the bottom left panel we show the bounds imposed by each one of the 11 {\sc Suzaku} fields. Now the kink occurs at $m_{\rm DM} \lesssim$ 10 MeV because the {\sc Suzaku} data are even lower in energy compared to {\sc NuStar} and {\sc Integral}. The fields (green in fig.~\ref{fig:galaxymap}) are all positioned at high latitudes and large longitudes and offer comparable bounds, with LL10 and LL21 slightly more stringent than the other ones. 

\medskip

Finally, in the bottom right panel we show the bounds imposed by {\sc Xmm-Newton} data considering each ring separately. We show for definiteness the data from the MOS camera (those from the PN camera turn out to be very similar but slightly less stringent). Each line/color in the plot corresponds to one $6^\circ$ degree ring as depicted in fig.~\ref{fig:galaxymap}. Not surprisingly, the inner rings (warmer colors in the figure) are more constraining because the DM density is higher in the inner galaxy. 
Note that the spread of the limits is wider for annihilating DM compared to decaying DM, as expected because of the different dependence of the source with the DM density ($\rho_{\rm DM}^2$ versus $\rho_{\rm DM}$, respectively).

\medskip

In figs.~\ref{fig:ResultsAnn} and \ref{fig:ResultsDec} we show the combined bounds for each experiment. This means that we apply the statistical criterion in eq.~(\ref{eq:chi2}) to the whole dataset of each experiment: the {\sc Integral} bounds are obtained using all the data of the 5 energy bands and the {\sc NuStar}, {\sc Suzaku} and {\sc Xmm-Newton} ones using all the regions of observation. 
The left panels refer to the DM (DM)$\to e^+e^-$ channel while the right panels to the DM (DM)$\to \mu^+\mu^-$ and DM (DM)$\to \pi^+\pi^-$ channels.
 
\medskip

\begin{figure}[t]
\begin{center}
\includegraphics[width= 0.80 \textwidth]{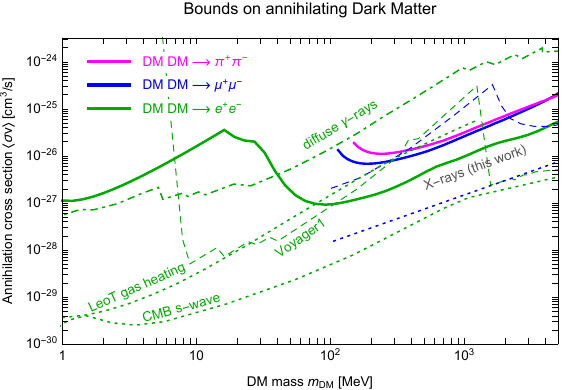} 
\caption{\em \small \label{fig:FinalResultsAnn} Final combined results for annihilating DM from this work compared with existing bounds. We report the bounds from Essig et al.~\cite{Essig:2013goa}, obtained using a compilation of $X$-ray and soft $\gamma$-ray data (dot-dashed green line marked `diffuse $\gamma$-rays'); the bounds from Boudaud et al.~\cite{Boudaud:2016mos} derived using data from {\sc Voyager 1} (dashed green and blue lines, corresponding to the $e^+e^-$ and $\mu^+\mu^-$ annihilation channels, respectively); the CMB bounds from Slatyer~\cite{Slatyer:2015jla} and Lopez-Honorez et al.~\cite{Lopez-Honorez:2013cua} (dotted green and blue lines, for the $e^+e^-$ and $\mu^+\mu^-$  channels); the bounds from gas heating in Leo T, obtained by Wadekar and Wang~\cite{Wadekar:2021qae} (also dotted, since the physics mechanism of energy injection is similar to the CMB one).}
\end{center}
\end{figure}

\begin{figure}[t]
\begin{center}
\includegraphics[width= 0.80 \textwidth]{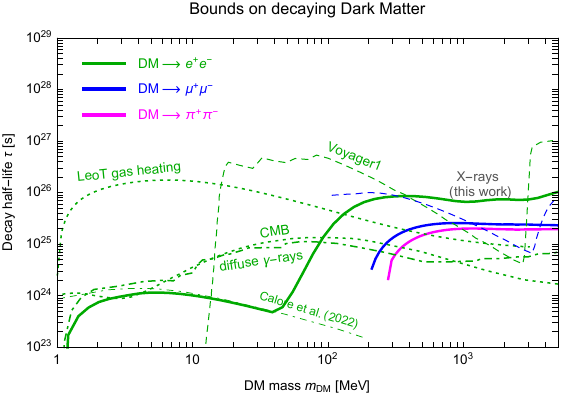} 
\caption{\em \small \label{fig:FinalResultsDec} Final combined results for decaying DM from this work compared with existing bounds. The constraints and the references are the same as in fig.~\ref{fig:FinalResultsAnn}, except that the CMB ones are derived in Liu et al.~\cite{Liu:2016cnk}. In addition, we plot the constraints of Calore et al.~\cite{Calore:2022pks}.}
\end{center}
\end{figure}

Figs.~\ref{fig:FinalResultsAnn} and \ref{fig:FinalResultsDec} represent our final results: we show only the most stringent constraints that we obtain, for the three annihilation/decay channels. 

For the case of DM annihilating into $e^+e^-$, {\sc Integral} imposes the bound $\langle \sigma v \rangle \lesssim 10^{-27}$ cm$^3$/s, around $m_{\rm DM} \simeq$ 100 MeV. 
The constraint loosens up to $\langle \sigma v \rangle \lesssim 10^{-25}$ cm$^3$/s around $m_{\rm DM} \simeq$ 20 MeV, where the bounds are imposed by {\sc NuStar}.
In the region $m_{\rm DM} \lesssim$ 10 MeV, the dominant contribution of the ICS component is too low in energy to be constrained by the data.
DM annihilating into $\mu^+\mu^-$ or $\pi^+\pi^-$ is constrained to $\langle \sigma v \rangle \lesssim 10^{-26} - 10^{-25}$ cm$^3$/s in the relevant mass interval.

For the case of DM decaying into $e^+e^-$, {\sc Integral} imposes the bound on the decay half-life $\tau = 1/\Gamma \gtrsim 10^{26}$ s, over the range $m_{\rm DM} \simeq$ 100 MeV $-$ few GeV. 

\begin{figure}[t]
\begin{center}
\includegraphics[width= 0.48 \textwidth]{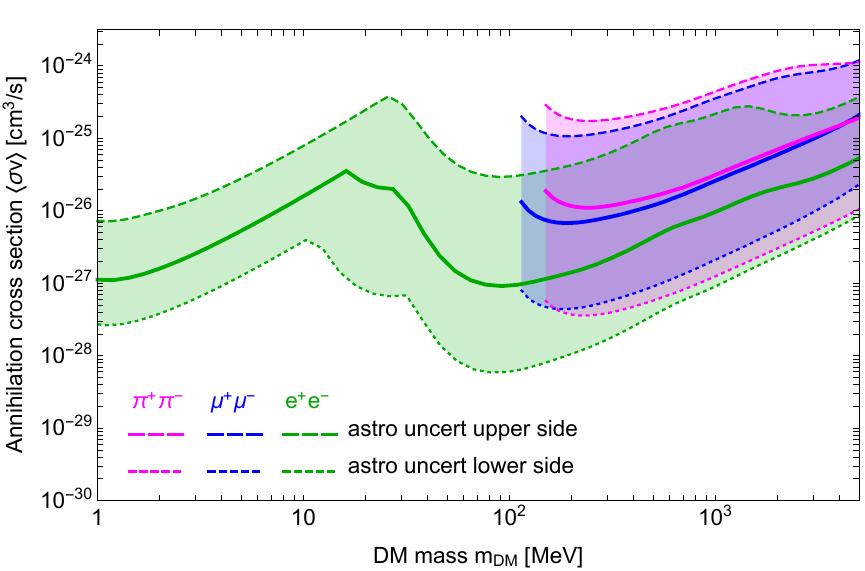} \quad
\includegraphics[width= 0.48 \textwidth]{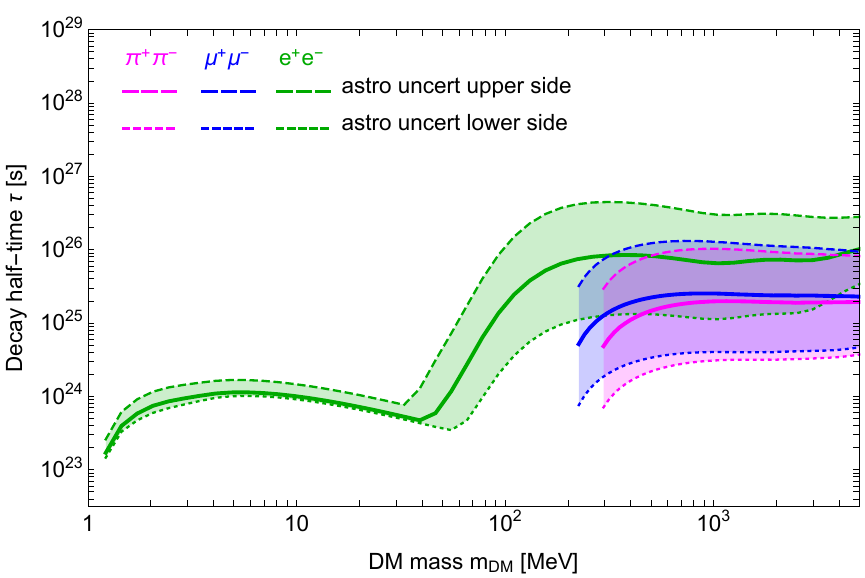} 
\caption{\em \small \label{fig:uncertainties} Illustration of the impact of astrophysical uncertainties on annihilation (left) and decay (right) constraints.}
\end{center}
\end{figure}

\smallskip

In fig.~\ref{fig:uncertainties} we show the impact of astrophysical uncertainties on annihilation (left) and decay (right) constraints, by following the same strategy as in \cite{Cirelli:2020bpc}. We vary the DM profile, the gas density in the Galaxy (which influences the energy losses by Coulomb interactions, ionization and bremsstrahlung), the radiation field density (affecting the energy losses but also ICS emission directly) and the galactic magnetic field. 
More precisely: we adopt a cored profile and a peaked NFW one (characterized by a slope $r^{1.26}$ towards the GC), we vary the gas and radiation field density by a factor of 2 above and below their central values, and we adopt the different configurations of the magnetic field discussed in \cite{Cirelli:2010xx}. We then compute the upper and lower envelopes of the $X$-ray fluxes from these combined variations, and we derive the corresponding bounds, resulting in the uncertainty bands of fig.~\ref{fig:uncertainties}. Note that the constraints can (generously) vary within two orders of magnitude.


\section{Comparison with related work}
\label{sec:comparison}
In this section we discuss how our results compare with the existing constraints in the literature.

Using a compilation of $X$-ray and soft $\gamma$-ray data from {\sc Heao-1}, {\sc Integral}, {\sc Comptel}, {\sc Egret} and {\sc Fermi}, Essig et al.~\cite{Essig:2013goa} have derived bounds on the $e^+e^-$ channel, shown as a dot-dashed line in fig.~\ref{fig:FinalResultsAnn}. This work does not include the ICS emission: indeed it leads to bounds that are comparable to ours in the small range where ICS is not relevant ($m_{\rm DM} \lesssim$ 10 MeV) and are instead weaker than ours at any larger mass where the ICS is the leading contribution to our limits. 

Using low energy measurements by {\sc Voyager 1} of the $e^\pm$ cosmic ray flux outside of the heliosphere, Boudaud et al.~\cite{Boudaud:2016mos} have derived constraints on the $e^+e^-$ and $\mu^+\mu^-$ channel, shown as dashed lines in fig.~\ref{fig:FinalResultsAnn}. We report the bounds of their propagation model B, characterized by weak reacceleration. Their constraints are stronger than ours only in a small mass interval around 10 MeV, for the $e^+e^-$ case. They are always weaker for the $\mu^+\mu^-$ case.

Using the impact on the CMB anisotropies of the $e^+e^-$ injection by DM annihilation events, Slatyer \cite{Slatyer:2015jla} and Lopez-Honorez et al.~\cite{Lopez-Honorez:2013cua} derived the constraints represented by the dotted lines in fig.~\ref{fig:FinalResultsAnn}. 
Our bounds are less stringent than these everywhere. However, as discussed in \cite{Cirelli:2020bpc}, the CMB constraints hold under the assumption that DM annihilation is speed-independent ($s$-wave). If the DM annihilation is $p$-wave, i.e.~$\langle \sigma v \rangle \propto v^2$, they weaken considerably. Our constraints are instead essentially insensitive to these difference \cite{Cirelli:2020bpc}, which implies that, for the $p$-wave scenario, our limits represent the most stringent bounds for $m_{\rm DM} \gtrsim$ 100 MeV.

Finally, Wadekar and Wang~\cite{Wadekar:2021qae} derived bounds from the requirement that DM $e^+e^-$ annihilations do not overheat the gas in the gas-rich dwarf galaxy Leo T. This constraint is represented by a dotted line in fig.~\ref{fig:FinalResultsAnn} and it is more stringent than ours for $m_{\rm DM} \lesssim 80$ MeV. However, this bound would relax significantly if the DM annihilation is $p$-wave (see \cite{Wadekar:2021qae} for details), similarly to the CMB constraints.

\medskip 

For the case of decaying DM, the existing constraints in the literature are shown in fig.~\ref{fig:FinalResultsDec}. 
The diffuse $\gamma$-ray constraints of Essig et al.~\cite{Essig:2013goa} are shown as a dot-dashed line, while the CMB and the dwarf gas heating constraints of \cite{Liu:2016cnk} and \cite{Wadekar:2021qae} respectively are shown as a dotted curve. 
The {\sc Voyager 1} constraints~\cite{Boudaud:2016mos} are dashed. Recently Calore et al.~\cite{Calore:2022pks} have considered the DM $\to e^+e^-$ channel (as well as the direct decaying channel DM $\to \gamma \gamma$, which is not of interest for us) and has used {\sc Integral/Spi} diffuse data, their bounds are displayed as a thin dot-dashed line. The constraints derived in this work (thick lines) are the most stringent limits for decaying DM in the range $m_{\rm DM} \simeq 400 \ {\rm MeV} - 3 \ {\rm GeV}$. Besides the $\mu^+\mu^-$ {\sc Voyager 1} ones, we are not aware of other existing constraints for the $\mu^+\mu^-$ and $\pi^+\pi^-$ channels in this mass interval.


\section{Conclusion}
\label{sec:conclusions}

In this paper we have focused on light, sub-GeV DM indirect detection, following up on the exploratory analysis performed in \cite{Cirelli:2020bpc}. DM in this mass range (1 MeV to $\sim$5 GeV) is notoriously difficult to probe with indirect searches, given the scarcity of MeV-GeV range experiments which could probe its soft $\gamma$-ray prompt emission. 
We have therefore concentrated on its secondary emission, which produces $X$-rays via the Inverse Compton Scattering of DM-produced $e^\pm$ over the galactic ambient light.
We have used data from the {\sc NuStar}, {\sc Suzaku}, {\sc Integral} and {\sc Xmm-Newton} satellites, in a number of different fields of observation in the Galaxy (see fig.~\ref{fig:galaxymap}). We have compared these measurements to the predicted flux from annihilating or decaying DM, considering the three relevant channels DM (DM) $\to e^+e^-, \mu^+\mu^-$ and $\pi^+\pi^-$. 

We find that the different $X$-ray surveys impose significant constraints. For decaying DM, they are the most stringent to date, in the range $m_{\rm DM} \simeq 400 \ {\rm MeV} - 3 \ {\rm GeV}$ (see fig.~\ref{fig:FinalResultsDec}). 
For annihilating DM, our limits are significant for $m_{\rm DM} \gtrsim$ 100 MeV, but the CMB $s$-wave bounds are still more stringent (see fig.~\ref{fig:FinalResultsAnn}). 
The sizeable astrophysical uncertainties related to the galactic DM distribution and the galactic environment can affect these results and make them tighter or looser by up to one order of magnitude in each direction (see fig.~\ref{fig:uncertainties}). 

\medskip

In the search for indirect detection of light DM, two possible avenues can be pursued, along the same lines of the discussion above. 
On one side, a few upcoming soft $\gamma$-ray missions such as {\sc as-/eAstrogam} (300 keV - 3 GeV)~\cite{DeAngelis:2017gra}, {\sc Amego} (200 keV - 10 GeV)~\cite{McEnery:2019tcm} or {\sc Cosi} (0.2 - 5 MeV)~\cite{Kierans:2017bmv,Tomsick:2021wed} will plug the `MeV gap' making it possible to directly probe the prompt emission of annihilating/decaying sub-GeV DM (see e.g.~\cite{Caputo:2022dkz}). 
On the other side, in the near future the full-sky data from the e{\sc Rosita} $X$-ray telescope (covering the range 0.2 - 10 keV)~\cite{eROSITA:2012lfj} will be available, as well as the radio surveys from the {\sc Ska} Observatory~\cite{Dutta:2020lqc}. Together with a better understanding of the galactic and dwarf-galaxy environments (ambient light, gas density and magnetic fields) and of the DM distribution, they will make it possible to leverage even more on the power of our indirect technique and thus improve its reach.

\medskip

\small
\subsubsection*{Acknowledgments}
\footnotesize
We are very grateful to Mitsuda Kazuhisa for providing us with the data from {\sc Suzaku} in digital form. 
We thank P.~De La Torre Luque, S.~Balaji, J.~Foster and C.~Dessert for discussions on the issue with the presentation of the data in Ref.~\cite{Foster:2021ngm}, which led to the revised bounds presented in this v3.
M.C.~and J.K.~acknowledge the hospitality of the Institut d'Astrophysique de Paris ({\sc Iap}) where part of this work was done. 
M.C. and N.F. acknowledge the hospitality of the CCA-Flatiron Institute and of New York University, where part of this work was done. \\[-3mm]

\noindent Funding and research infrastructure acknowledgements: 
{\sc Cnrs} grant {\sl DaCo:~Dark Connections}, in collaboration with Alma Mater Studiorum - Universit\`a di Bologna [work of M.C. and J.K.];  {\sc Departments of Excellence} grant awarded by the Italian Ministry of Education, University and Research ({\sc Miur})  [work of N.F. and E.P.]; Research grant {\sl The Dark Universe: A Synergic Multimessenger Approach}, No. 2017X7X85K funded by the Italian Ministry of Education, University and Research ({\sc Miur})  [work of N.F. and E.P.]; Research grant {\sc TAsP} (Theoretical Astroparticle Physics) funded by Istituto Nazionale di Fisica Nucleare ({\sc Infn})  [work of N.F. and E.P.]; {\sl Fermi Research Alliance, LLC} under Contract No. DE-AC02-07CH11359 with the U.S. Department of Energy, Office of High Energy Physics.  [work of E.P.].

\bibliographystyle{JHEP-noquotes}
\bibliography{XrayDM2.bib}

\end{document}